\documentclass[12pt]{article} % Specifies the document class
\usepackage{epsf}%
\usepackage{latexsym}
\usepackage[dvips]{graphicx}
\usepackage{mathrsfs}

\newcommand {\be}{\begin{equation}}
\newcommand {\ee}{\end{equation}}
\newcommand{\bea}{\begin{eqnarray}}
\newcommand{\eea}{\end{eqnarray}}
\newcommand{\ba}{\begin{array}}
\newcommand{\ea}{\end{array}}
\newcommand{\beq}{\begin{eqnarray*}}
\newcommand{\eeq}{\end{eqnarray*}}
\newcommand{\ds}{\displaystyle}

\newcommand{\AF}{{\rm AF}}
\newcommand{\NP}{{\rm NP}}
\newcommand{\MC}{{\rm MC}}

\def\L{{L}}

\begin{document}

\begin{center}
{\Large \bf Peculiar Properties of SU(2) Gauge Field Thermodynamics
on a Finite Lattice. Calculation of Beta-function  }
\bigskip\bigskip

{\large O.A. Mogilevsky}

\bigskip
{\it Bogolyubov Institute for Theoretical Physics}\\
{\it (14b, Metrolohichna Str., Ky\"\i v 03143, Ukraine}\\
{\it e-mail: mogil@bitp.kiev.ua}
\end{center}
\bigskip

{\large

\begin{center}
{\bf Abstract}\\
\medskip
The new method of nonperturbative calculation of the beta-function
in the lattice gauge theory is proposed. The method is based on the
finite size scaling hypothesis.
\end{center}

Ever since the pioneering work by Creutz [1] the approach to
asymptotic scaling, and thus the continuum limit, was one of the
central issues in studies of gauge theories on the lattice. Although
the first results were promising, the lack of asymptotic  scaling of
physical observables has been observed in SU(N) gauge theories. One
of the main source of the nonperturbative results in the gauge
theories today is the Monte-Carlo (MC) lattice calculations. For the
SU(N) pure gauge theories on lattices of size $N_{\tau}\times
N_{\sigma}^{3}$ MC results are the dimensionless functions of the
bare coupling constant $g$ (another form for the coupling,
$\beta\equiv 2N/g^{2}$, is often used). The transformation of these
functions to physical quantities are done by multiplying them on
lattice spacing $a$ in the corresponding powers. The length scale
$\L$ and the temperature  $T$ are given as

\be \L=N_{\sigma} a, \qquad T=\left(N_{\tau}a\right)^{-1}. \ee

To define the physical quantities one needs a connection between
lattice spacing $a$ and bare coupling constant $g$. Such a
connection is formulated in terms of the beta-function
$\beta_{f}(g)$ through the equation

\be \beta_{f}(g) = -a{dg\over da}. \ee

\noindent The perturbation theory gives the asymptotic expansion of
the beta-function

\bea \beta_{f}^{\AF} &=& - b_{0}g^{3} -b_{1}g^{5}+O(g^{7}), \nonumber\\[0.3cm]
b_{0}&=& {11N\over 48\pi^{2}}, \quad b_{1}={34\over3}\left( {N\over 16\pi^{2}}
\right)^{2},
\eea

\noindent where $N=2$ in the SU(2) case. The differential equation
(2) with $\beta_{f}^{\AF}(g)$ in (3) leads to

\be a\Lambda_{\L}^{\AF} =\exp \left(-{1\over 2b_{0}g^{2}}\right)
\cdot \left(b_{0}g^{2}\right)^{-b_{1}/2b_{0}^{2}} \equiv R(g^{2}),
\ee

\noindent where $\Lambda_{\L}^{\AF}$ is the renormalization group
invariant parameter (integration constant of Eq.~(2)). Eq.~(4) is
known as asymptotic freedom (AF) relation.

Using (1) and (4) one can calculate

\be {T_{c}\over \Lambda_{\L}^{\AF}} ={1\over N_{\tau}R(g_{c}^{2})}.
\ee

\noindent The values of $T_{c}/\Lambda_{\L}^{\AF}$ at different
$N_{\tau}$ are presented in Table 1. One observes a rather strong
dependence of $T_{c}/\Lambda_{\L}^{\AF}$ on $N_{\tau}$. This means
that the perturbative AF relation (4) does not work even on the
largest available lattices. This fact is known as absence of the
asymptotic scaling.

It has been proposed in Ref. [3] that a deviation from the
asymptotic scaling can be described by a universal non-perturbative
(NP) beta-function, i.e. $\beta_{f}(g)$ is the same one for all
lattice observables and it does not depend on the lattice size if
$N_{\sigma}$ and $N_{\tau}$ are not too small.

The following ansatz was suggested [3]:

\be a\Lambda_{\L}^{\NP} =\lambda (g^{2}) R(g^{2}), \ee

\noindent where $R(g^{2})$ is given by (4) and $\lambda(g^{2})$ is
thought to describe a deviation from perturbative behaviour. The
equation (4) has been expected at $g\rightarrow 0$ so that an
additional constraint, $\lambda(0)=1$, has been assumed. The values
of $T_{c}/\Lambda_{\L}^{\NP}$ can be calculated then as

\be T_{c}/\Lambda_{\L}^{\NP} = {1\over N_{\tau} \lambda(g_{c}^{2})
R(g_{c}^{2})}. \ee

A simple formula for the function $\lambda(g^{2})$ was suggested
[3]:

\be \lambda(g^{2}) =\exp \left( {c_{3}g^{6}\over 2b_{0}^{2}}\right).
\ee

\noindent Parameter $c_{3}$ in (8) and a new one,
$T_{c}^{*}/\Lambda_{\L}^{\NP}$=const, have been considered as free
parameters and determined from fitting the MC values of
$T_{c}/\Lambda_{\L}^{\NP}$ (7) at different $N_{\tau}$ to the
constant value $T_{c}^{*}/\Lambda_{\L}^{\NP}$. This procedure gives

\be T_{c}^{*}/\Lambda_{\L}^{\NP} =21,45(14), \quad
c_{3}=5,529(63)\cdot 10^{-4}. \ee

\noindent In comparison to $T_{c}/\Lambda_{\L}^{\AF}$ the much
weaker $N_{\tau}$ dependent values of $T_{c}/\Lambda_{\L}^{\NP}$
have been obtained, which become now close to the constant value
$T_{c}^{*}/\Lambda_{\L}^{\NP}$ (9).

Despite of the phenomenological success of the above procedure of
[3] the crucial question, regarding the existence of the universal
NP beta-function which does not depend on the lattice size, is not
solved and remains just a postulate. A principal difference of our
approach is that we do not assume the existence of the universal
beta-function and take into account the finite size effects of the
lattice.

Usually finite size scaling (FSS) in the vicinity of a
finite-temperature phase transition is discussed for lattice SU(N)
gauge models without trying to make contact with the continuum
limit, i.e. the scaling properties are studied on lattices
$N_{\tau}\times N_{\sigma}^{3}$ with fixed $N_{\tau}$ and varying
$N_{\sigma}$, and the model is viewed as a three-dimensional spin
system. On the other hand, in the continuum limit the FSS properties
of these non-abelian models should, of course, be discussed in terms
of the physical volume $V=L^{3}$ and the temperature $T$. On a
$N_{\tau}\times N_{\sigma}^{3}$ lattice $L$ and $T$ are given in
units of the lattice spacing $a$, therefore it is advantageous to
introduce the dimensionless combination

\be LT={N_{\sigma}\over N_{\tau}}. \ee

\noindent The scaling behaviour of the continuum theory emerges from
the lattice free energy on arbitrary lattices, i.e. when varying
$N_{\tau}$ and $N_{\sigma}$.

Following [2] let us discuss briefly the FSS procedure. The singular
part of the free energy density is described by the universal
finite-size scaling function

\be f \left(t,h,N_{\sigma}, N_{\tau}\right)
= \left({N_{\sigma}\over N_{\tau}}\right)^{-3} Q_{f}
\Biggl( g_{t}\left({N_{\sigma}\over N_{\tau}}\right)^{1\over \nu}, g_{h}
\left({N_{\sigma}\over N_{\tau}}\right)^{\beta+\gamma\over \nu}\Biggr),
\ee

\noindent where $\beta, \gamma, \nu$ are the critical indexes of the
theory, the scaling function $Q_{f}$ depends on the reduced
temperature $t=(T-T_{c})/T_{c}$ and the external field strength $h$
through thermal and magnetic scaling fields

\bea
g_{t} &=& c_{t}t\left( 1+b_{t}t\right)\\[0.3cm]
g_{h} &=& c_{h}h\left(1+b_{h}t\right)
\eea

\noindent with non-universal coefficients $c_{t}, c_{h}, b_{t},
b_{h}$ still carrying a possible $N_{\tau}$ dependence.

The order parameter and the susceptibility are now obtained as
derivatives of $f$

\bea
\langle \L\rangle &=& -{\partial f\over \partial h}\Biggl|_{h=0}
= \left({N_{\sigma}\over N_{\tau}}\right)^{-\beta/\nu} Q_{\L} \Biggl(
g_{t} \left({N_{\sigma}\over N_{\tau}}\right)^{1/\nu}\Biggr)\\[0.3cm]
\chi &=& {\partial f^{2}\over \partial h^{2}}\Biggl|_{h=0}
= \left({N_{\sigma}\over N_{\tau}}\right)^{\gamma/\nu} Q_{\chi}
\Biggl(g_{t} \left({N_{\sigma}\over N_{\tau}}\right)^{1/\nu}\Biggr)
\eea

\noindent Here we have used the hyperscaling relation

\beq
{\gamma\over \nu} +2{\beta\over\nu}=3
\eeq

\noindent Taking the fourth derivative of $f$ at $h=0$ it is easy to
see that the quantity

\be g_{4} ={\partial^{4}f\over \partial h^{4}}\Biggl|_{h=0} \Biggl/
\chi^{2} \left({N_{\sigma}\over N_{\tau}}\right)^{3} \ee

\noindent is directly a scaling function

\be g_{4}=Q_{g_{4}}\Biggl( g_{t}\left({N_{\sigma}\over
N_{\tau}}\right)^{1/\nu} \Biggr). \ee

\noindent On a finite lattice $g_{4}$ has the form

\be g_{4} ={\langle \L^{4}\rangle\over \langle \L^{2}\rangle^{2}}-3,
\ee

\noindent i.e. it is the normalized fourth cumulant of the Polyakov
loop.

Our approach is based on the two points: i) more conventional
statistical mechanical definition of the beta-function and ii) FSS
and the phenomenological renormalization. Let us make the
infinitesimal transformation of the lattice spacing $a\rightarrow a'
=ba=(1+\Delta b)a$. Then

\be -a{dg\over da} =-\lim\limits_{b\rightarrow1} \left(
a{g(ba)-g(a)\over ba-a}\right) =-\lim\limits_{b\rightarrow1}
{dg\over db}. \ee

\noindent We get the definition of the beta-function for the lattice
system%%%
\be \beta_{f}(g) =-\lim\limits_{b\rightarrow1} {dg\over db}. \ee

Let us first consider the case when $N_{\tau}$ is kept a fixed one.
Then $N_{\tau}$ can be absorbed in the non-universal constants in
$g_{t}$ and $g_{h}$ and we deal with the usual form at the FSS as in
the standard spin theory (see, for example [4]). The existence of
the scaling function $Q$ allows  to develop a procedure to
renormalize the coupling constant $g^{-2}$ by using two different
lattice sizes $N_{\sigma}$ and $N'_{\sigma}$. Let us fix the spatial
size
$\L=N_{\sigma}a$ and make a scale transformation %%%
\bea a& \rightarrow & a'=ba\nonumber\\
N_{\sigma} &\rightarrow & N'_{\sigma}=N_{\sigma}/b. \eea

\noindent Then the phenomenological renormalization is defined by
the following equation %%%
\be Q(g^{-2}, N_{\sigma}) = Q\Bigl( (g')^{-2}, N_{\sigma}/b\Bigr).
\ee

\noindent It expresses that the scaling function $Q$ remains  to be
unchanged if the lattice size is rescaled by a factor $b$ and the
inverse coupling $g^{-2}$ is shifted to $(g')^{-2}$ simultaneously.
Taking the derivative with respect to the scale parameter $b$ of the
both sides of (22) and using (20) it is easy to obtain the
expression

\be a{dg^{-2}\over da} ={\partial\ln Q/\partial\ln N_{\sigma}\over
\partial \ln Q/\partial g^{-2}}.
\ee

\noindent The approximation of the derivative with respect to
$N_{\sigma}$ by the finite difference yields the formula for
beta-function

\be a{dg^{-2}\over da} ={\ln {\ds Q(N'_{\sigma})\over\ds
Q(N_{\sigma})} / \ln ({\ds N'_{\sigma}\over \ds N_{\sigma}})\over
\left[ {\ds dQ(N_{\sigma})\over\ds dg^{-2}}\cdot {\ds
dQ(N'_{\sigma})\over\ds dg^{-2}}/ Q(N_{\sigma})
Q(N'_{\sigma})\right]^{1/2}}\, . \ee

\noindent Further we consider the formula (24) for the fourth
cumulant $g_{4}(g^{-2}, N_{\sigma})$, which is the scaling function
directly. Fig. 1 presents the MC data for $g_{4}$ on the lattices
$N_{\tau}=4$; $N_{\sigma}=12, 18, 26, 36$ [5]. One can easy to see
from (24) that beta-function has a zero at the fixed point
$4/g_{c}^{2}=2,299$ of the renormalization transformation (22) in
full accordance with a second-order nature of the deconfinement
phase transition in SU(2) lattice gauge theory.

\begin{figure}[ph]\centering
  \includegraphics[width=10.cm]{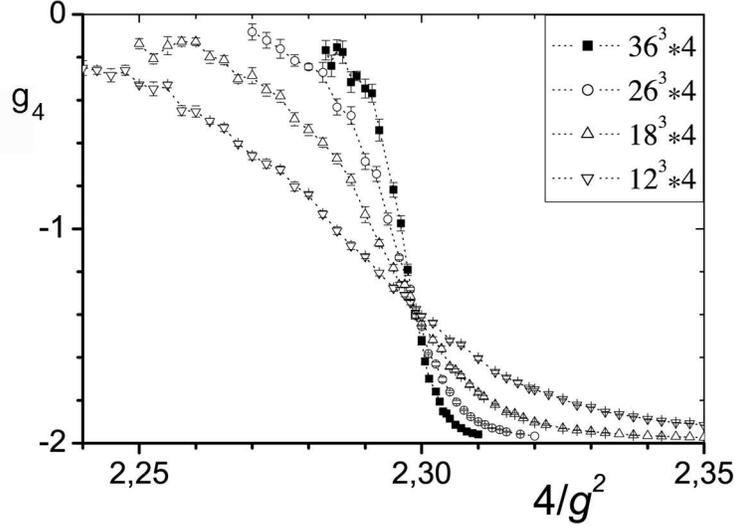}\\
  \caption{The fourth Binder cumulant $g_{4}$ on the lattices $N_{\tau}=4$;
  $N_{\sigma}=12,18,26,36$. MC data are taken from [5].}
\end{figure}

The results of the calculation of beta-function according to formula
(24) are presented in Fig. 2 for three sets $N_{\sigma}=12$,
$N'_{\sigma}=18$; $N_{\sigma}=18$, $N'_{\sigma}=26$;
$N_{\sigma}=26$, $N'_{\sigma}=36$. Although $N_{\sigma}$ and
$N'_{\sigma}$ in the different pairs are not too close, one can see
surprisingly the coincidence of the curves at $4/g^{2}\geq
4/g_{c}^{2}$. This observation gives the hope that beta-function
does not depend on the spatial size of lattice in the deconfinement
phase.

\begin{figure}[ph]%\centering
~~~~~~~~~~ \includegraphics[width=11.8cm]{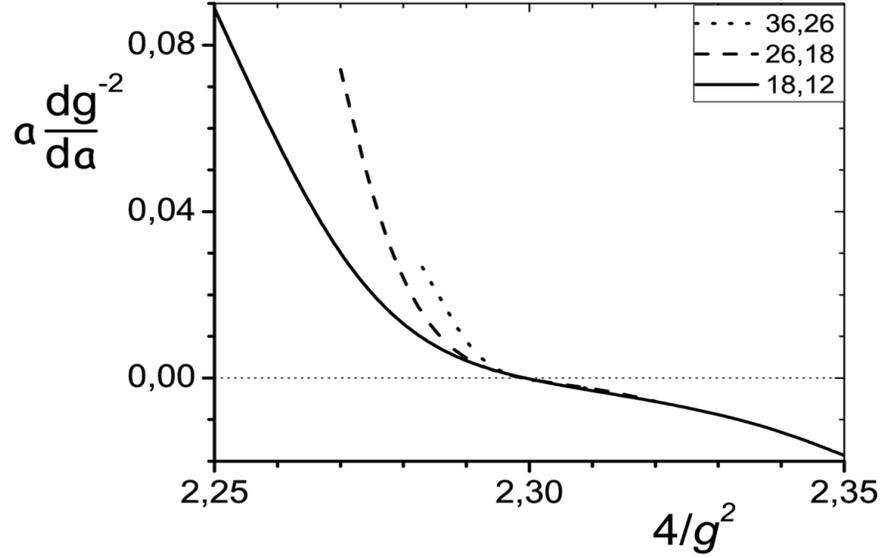}\\
  \caption{Beta-function from (24) for the pairs
  $N_{\sigma}=12$,
$N'_{\sigma}=18$; $N_{\sigma}=18$, $N'_{\sigma}=26$;
$N_{\sigma}=26$, $N'_{\sigma}=36$.}
\end{figure}

Next we consider fixed $y=N_{\sigma}/N_{\tau}$, by varying
$N_{\sigma}$, and therefore $N_{\tau}$ accordingly as is needed to
reach the continuum limit. Rescaling $N_{\sigma}$ and $N_{\tau}$ by
a factor $b$ leads to a phenomenological renormalization by the
following  identity for a scaling function $Q$ %%%
\be Q\Biggl( g_{t}\left(g^{-2}, N_{\tau}\right)\cdot \left(
{N_{\sigma}\over N_{\tau}}\right)^{1/\nu}\Biggr) = Q \Biggl(
g_{t}\left( (g')^{-2},N_{\tau}/b\right)\cdot \left(
{bN_{\sigma}\over bN_{\tau}}\right)^{1/\nu}\Biggr), \ee

\noindent where $g_{t}(g^{-2}, N_{\sigma})$ is determined by (12).
If we ignore the possible $N_{\tau}$ dependence of the coefficients
$c_{t}$ and $b_{t}$, then it follows from (25)

\be t\left( g^{-2}, N_{\tau}\right) =t\left( (g')^{-2},
N_{\tau}/b\right). \ee

\noindent In general the reduced temperature $t=(T-T_{c})/T_{c}$ is
a complicated function of the coupling $\beta=2N/g^{2}$, which in
the vicinity of the critical temperature $T_{c}$ can be approximated
by [2]

\be t=(\beta-\beta_{c}) {1\over 4Nb_{0}} \left[ 1-{2Nb_{1}\over
b_{0}} \beta_{c}^{-1}\right]. \ee

\noindent This approximation reproduces the correct reduced
temperature in the continuum limit, which is easy verified by using
(4). Taking the derivatives with respect to the scale parameter $b$
of the both sides of (26) and using (20) and (27) it is easy to
obtain the expression for the beta-function

\be \beta_{f}(g) =-B_{0}(N_{\tau}) g^{3} -B_{1}(N_{\tau})g^{5}, \ee

\noindent where

\be \left\{ \ba{l} B_{0}(N_{\tau}) ={\ds 1\over\ds 4N} \left( 1-{\ds
2Nb_{1}\over \ds b_{0}\beta_{c}}\right) {\ds d\beta_{c}\over\ds d\ln
N_{\tau}}\\[0.3cm]
B_{1}(N_{\tau}) = B_{0}(N_{\tau}) {\ds b_{1}\over \ds b_{0}}.\ea
\right. \ee

\noindent Then the equation (2) leads to %%%
\be a\Lambda_{\L} =\exp \left(-{1\over 2B_{0}g^{2}}\right)
\left(B_{0}g^{2}\right)^{-B_{1}/2B_{0}^{2}}. \ee

\noindent Using (1) one can obtain the critical temperature $T_{c}$.
The problem only remains to calculate the derivative
$d\beta_{c}/d\ln N_{\tau}$ in expression (29). The calculation has
been made for the SU(2) gauge theory by fitting the MC data for
critical couplings $\beta_{c}^{\MC}(N_{\tau})$ with a spline
interpolation and numerical differentiation of this curve. The
result of the calculation is presented in Table 1. In comparison to
$T_{c}/\lambda_{\L}^{\AF}$ the much weaker dependence on $N_{\tau}$
of the critical temperature $T_{c}/\Lambda_{\L}$ is observed.

\begin{table}[h]
\centering
\begin{tabular}{|c|c|c|c|c|}
  \hline
&&&&\\
  $N_{\tau}$ &$\beta_{c}=4/g_{c}^{2}$ &
  $T_{c}/\Lambda_{\L}^{\AF}$ &  $d\beta_{c}/dN_{\tau}$ &
$T_{c}/\Lambda_{\L}$ \\
&&&&\\
  \hline
  &&&&\\
2  & 1.880 & 29.7 &--  &--\\
 3 & 2.177 & 41.4 & 0.158 & 25.22 \\
  4 & 2.299 & 42.1 & 0.086 & 25.46 \\
  5 & 2.373 &40.6& 0.063 & 25.38 \\
  6 & 2.427 & 38.7 & 0.045 &24.13 \\
  8 & 2.512 & 36.0 & 0.040 & 24.24 \\
  16 & 2.739 & 32.0 & 0.017 & -- \\ &&&&\\
  \hline
\end{tabular}

  \caption{MC data for $\beta_{c}$ are taken from [2]. The values of
  $T_{c}/\Lambda_{\L}^{\AF}$ are calculated from (4).
  Our results for $T_{c}/\Lambda_{\L}$ are
  obtained from (30).}
\end{table}

{}

}

\end{document}